\documentclass[twocolumn,secnumarabic, showpacs, superscriptaddress,amssymb,english,aps, prb,floatfix, longbibliography]{revtex4-1}

\usepackage{graphicx}
\usepackage{SIunits}
\usepackage{epstopdf}

\begin{document}


\title{Disentangling nonradiative recombination processes in Ge micro-crystals on Si substrates} 



\author{Fabio Pezzoli}
\email[]{fabio.pezzoli@unimib.it}
\affiliation{LNESS and Dipartimento di Scienza dei Materiali, Universit\`a  degli Studi di Milano-Bicocca, via Cozzi 55, I-20125 Milano, Italy}

\author{Anna Giorgioni}
\affiliation{LNESS and Dipartimento di Scienza dei Materiali, Universit\`a  degli Studi di Milano-Bicocca, via Cozzi 55, I-20125 Milano, Italy}

\author{Kevin Gallacher}
\affiliation{University of Glasgow, School of Engineering, Rankine Building, Oakfield Avenue, Glasgow G12 8LT,
United Kingdom}

\author{Fabio Isa}
\affiliation{LNESS, Dipartimento di Fisica del Politecnico di Milano and IFN-CNR, Polo Territoriale di Como, Via Anzani 42, I-22100 Como, Italy}
\affiliation{Laboratory for Solid State Physics, ETH Zurich, Otto-Stern-Weg 1, CH-8093 Z\"{u}rich, Switzerland}

\author{Paolo Biagioni}
\affiliation{LNESS, Dipartimento di Fisica del Politecnico di Milano and IFN-CNR, P.zza Leonardo da Vinci 32, I-20133 Milano, Italy}

\author{Ross W. Millar}
\affiliation{University of Glasgow, School of Engineering, Rankine Building, Oakfield Avenue, Glasgow G12 8LT,
United Kingdom}

\author{Eleonora Gatti}
\affiliation{LNESS and Dipartimento di Scienza dei Materiali, Universit\`a  degli Studi di Milano-Bicocca, via Cozzi 55, I-20125 Milano, Italy}

\author{Emanuele Grilli}
\affiliation{LNESS and Dipartimento di Scienza dei Materiali, Universit\`a  degli Studi di Milano-Bicocca, via Cozzi 55, I-20125 Milano, Italy}

\author{Emiliano Bonera}
\affiliation{LNESS and Dipartimento di Scienza dei Materiali, Universit\`a  degli Studi di Milano-Bicocca, via Cozzi 55, I-20125 Milano, Italy}

\author{Giovanni Isella}
\affiliation{LNESS, Dipartimento di Fisica del Politecnico di Milano and IFN-CNR, Polo Territoriale di Como, Via Anzani 42, I-22100 Como, Italy}

\author{Douglas J. Paul}
\affiliation{University of Glasgow, School of Engineering, Rankine Building, Oakfield Avenue, Glasgow G12 8LT, United Kingdom}

\author{Leo Miglio}
\affiliation{LNESS and Dipartimento di Scienza dei Materiali, Universit\`a  degli Studi di Milano-Bicocca, via Cozzi 55, I-20125 Milano, Italy}



\begin{abstract}
We address nonradiative recombination pathways by leveraging surface passivation and dislocation management in $\mu$m-scale arrays of Ge crystals grown on deeply patterned Si substrates. The time decay photoluminescence (PL) at cryogenic temperatures discloses carrier lifetimes approaching 45 ns in band-gap engineered Ge micro-crystals. This investigation provides compelling information about the competitive interplay between the radiative band-edge transitions and the trapping of carriers by dislocations and free surfaces. Furthermore, an in-depth analysis of the temperature dependence of the PL, combined with capacitance data and finite difference time domain modeling, demonstrates the effectiveness of $\mathrm{GeO_2}$ in passivating the surface of Ge and thus in enhancing the room temperature PL emission.
\end{abstract}


\maketitle 

The monolithic development of Si photonic components addresses the hurdles of charged-based computing and is expected to drastically transform information and communication technologies. The progress of Si photonics, however, has been jeopardized by fundamental limitations. The centrosymmetric crystal structure and the indirect nature of the band-gap of bulk Si imply weak electro-optical effects alongside poor absorption and emission efficiencies.\cite{Leuthold10} These obstacles have stimulated the exploration of wafer-scale fabrication methods for utilizing materials better suited for high-performance photonic circuitry.\cite{Dai12, Lim14} In this respect, direct epitaxial growth of Ge on Si substrates is a sought-after solution because Ge, being non-polar and isoelectronic with Si, turns out to be compatible with conventional microelectronic processes yielding high-volume throughput.\cite{Pillarisetty11} 

Another crucial benefit is that Ge, an indirect band-gap semiconductor, offers an unexpectedly strong light-matter interaction.\cite{Pezzoli13} The close energy proximity of $\approx 140$ meV between the fundamental conduction band minimum at the $L$ point of the Brillouin zone and the local minimum at the zone center $\Gamma$ guarantees optical access to the direct band-gap properties: a prerequisite for the fabrication of high performance photonic devices.\cite{Michel10, Liang10, Baldassarre15} Moreover, heteroepitaxial growth encompasses strain and alloying.\cite{Paul04} These two degrees of freedom were leveraged for decreasing the $\Gamma$-to-$L$ offset in the recent demonstrations of tensile-Ge \cite{Liu10,Camacho12} and GeSn\cite{Wirths15} injection lasers on Si, eventually turning Ge even closer to a direct band gap material. All these efforts proved Ge heteroepitaxy to be viable, but at the same time heralded very poor emission efficiencies. Epitaxial growth of Ge on Si faces severe material and manufacturing issues due to the large mismatch in the lattice constants and the thermal expansion coefficients. The former leads to nucleation of dislocations, while the latter causes wafer bowing and cracking. \cite{Paul04, Yang03} As a consequence, there is an urgent need to find reliable solutions to the ubiquitous nonradiative recombination of carriers at such growth defects and free surfaces, which considerably compromise the electronic and optical properties.\cite{Paul04, Sukhdeo16, Li16} 

\begin{figure*}
\includegraphics[width=13cm]{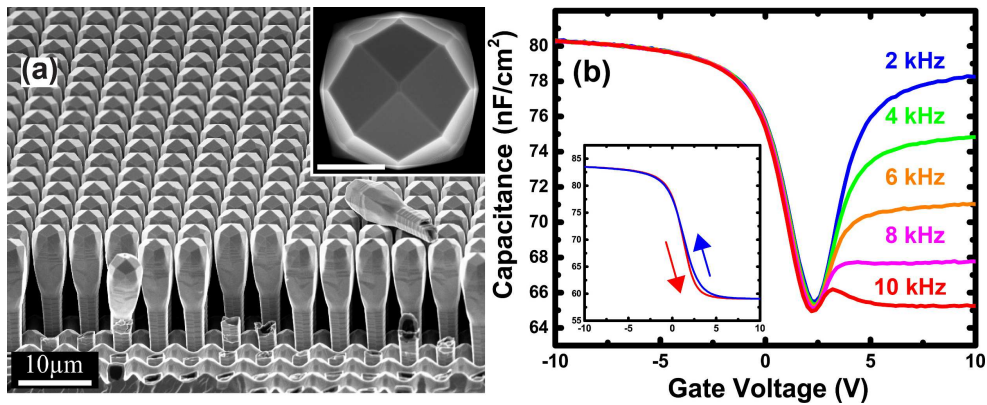}
 \caption{(a) A cross-sectional scanning electron microscope image of the as-grown $\mu$m-scale crystals developed by depositing 8 $\mu$m of Ge onto $\mathrm{2\times 2 \mu m^2}$ Si pillars. (b) The room temperature capacitance versus voltage (C-V) characteristics as a function of low frequency (2-10 kHz) for a Pt/$\mathrm{Si_3 N_4}$/$\mathrm{GeO_2}$/Ge capacitor. The inset shows the room temperature high frequency (1 MHz) bidirectional C-V curves.}
	\label{fig1}
 \end{figure*}

In the quest to achieve Ge-based architectures as loss-less optical components and to resolve the subtleties of the carrier dynamics, we exploit the out-of-equilibrium growth of Ge on deeply patterned Si substrates.\cite{Falub2012} Figure~\ref{fig1}(a) demonstrates a scanning electron microscope micrograph of as-grown $\mu$m-scale crystals developed by depositing 8 $\mu$m of Ge at $550 \,^{\circ}{\rm C}$ by low-energy plasma enhanced chemical vapor deposition\cite{Rosenblad98} onto $2\times2$ $\mathrm{\mu m^2}$ Si pillars. Such pedestals were separated by 8-$\mu$m-deep and 3-$\mu$m-wide trenches patterned onto (001) substrates by optical lithography and reactive ion etching.\cite{Falub2012} The Ge micro-crystals are characterized by a pyramidal top surface bounded by $\{113\}$ and $\{111\}$ facets. Such facets stem from a slanted growth front that promptly steers the threading arms towards the lateral sidewalls, where they become pinned. Subsequent material deposition eventually yields a region of the crystal that is completely free from dislocations, although no annealing was performed.\cite{Falub2012, Marzegalli13, Isa13} Such an approach has been shown to produce space-filling arrays of Ge featuring (i) a giant enhancement in the internal quantum efficiency caused by the removal of dislocations threading through the whole epitaxial layer and (ii) an improved light extraction enabled by total internal reflection at the sidewalls of the micro-crystals. \cite{Pezzoli14, Isa15}  

In this letter we utilize dislocation management provided by the Ge micro-crystals while passivating surface traps by means of conformal deposition of dielectric layers. By doing so, we can control and modulate the weight of the nonradiative recombination pathways and gather selective access to the carrier loss occurring either at extended defects or surface states in various temperature regimes. Aside from fruitful applications utilizing Ge as a photonic building block, pinpointing the parasitic sinks of charge carriers provides an in-depth knowledge of the physics of the recombination dynamics at the semiconductor interfaces, which is crucial also as a guide to design advanced semiconductor heterostructures that fulfill ever-demanding photonic and electronic functions.

\begin{figure}
\includegraphics[width=8cm]{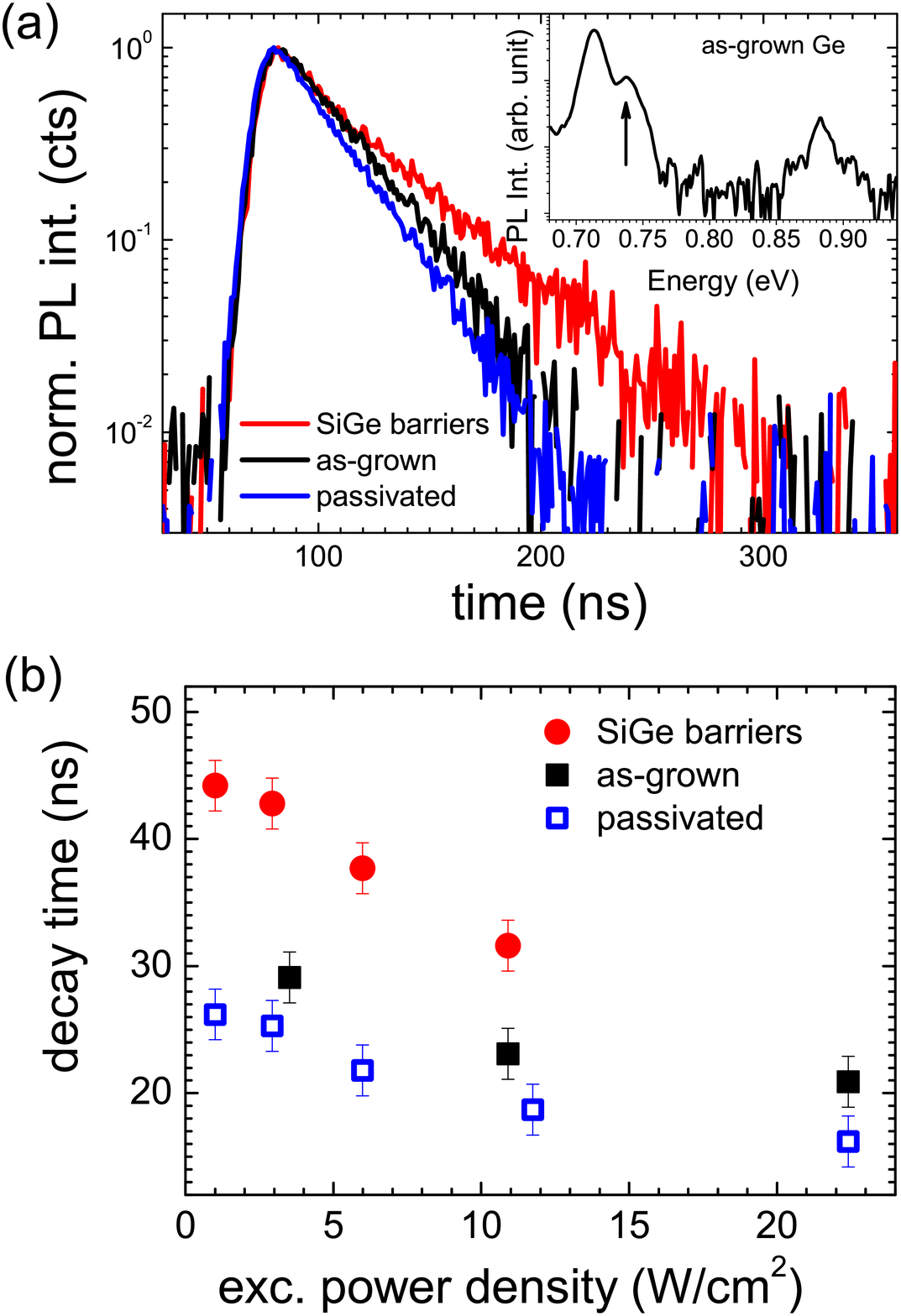}
 \caption{(a) The PL decay curves measured at a temperature of 14 K and an average excitation power density of about 3 W/$\mathrm{cm^2}$ for the as-grown (black line) Ge micro-crystals and for the micro-crystals either passivated with the oxide coating (blue line) or embedding three SiGe barriers (red line). The inset shows a low temperature continuous-wave PL spectrum for the as-grown sample and the arrow points to the energy of the indirect band-gap peak used for the PL decay measurements. (b) The power dependence of the electron lifetime extracted from the PL decay data for the three samples: as-grown (black squares), passivated (blue open squares) and micro-crystals blending SiGe barriers (red dots).}
	\label{fig2}
 \end{figure}

Surface passivation of the Ge micro-crystals was obtained by growing a high quality thermal $\mathrm{GeO_2}$ layer. \cite{Matsubara08} The Ge micro-crystal surfaces were first cleaned in acetone followed by rinsing in isopropanol before cyclic hydrofluoric acid / deionized water etching was performed to remove any native oxide films. The Ge micro-crystals were then immediately transferred to a furnace tube oven for thermal oxidation at $550 \,^{\circ}{\rm C}$ with $\mathrm{O_2}$ gas for 60 min, which results in a $\mathrm{GeO_2}$ thickness of about 20 nm. The $\mathrm{GeO_2}$ was then subsequently capped with 50 nm of unstrained ICP-PECVD $\mathrm{Si_3 N_4}$.\cite{Millar15} 

To evaluate the quality of the thermally grown $\mathrm{GeO_2}$, capacitor structures consisting of $\mathrm{Pt/Si_3 N_4 / GeO_2/Ge}$ were fabricated in parallel. Figure~\ref{fig1}(b) demonstrates the room temperature capacitance-voltage (C-V) characteristics of the $\mathrm{GeO_2}$ capacitor structures at low frequencies (2-10 kHz). The well behaved threshold voltage response as a function of frequency indicates that there is a low level of interface trapped charge between the $\mathrm{GeO_2}$ and Ge. The inset in Fig.~\ref{fig1}(b) shows the high frequency (1 MHz) bidirectional C-V curves. There is no flat-band voltage shift observed from the ideal that was calculated from the work function of Pt ($\approx 5.6$ eV), suggesting there is negligible fixed charge present within the dielectric stack. There is a relatively small flatband voltage hysteresis observed ($\approx 300$ mV), which is still consistent with high quality $\mathrm{GeO_2}$/Ge interfaces.\cite{Bellenger08, Hirayama11} All these observations indicate an efficient electrical passivation of the dangling bonds on the surface of Ge.

In addition to the C-V investigation, we have also carried out photoluminescence (PL) decay measurements of the indirect band-gap transition to gather direct insights about the effectiveness of the surface passivation on the optical properties of the Ge micro-crystals. The exceedingly long radiative lifetime expected in an indirect gap material such as Ge is typically concealed by the presence of competitive nonradiative events occurring at the defect sites, which appreciably shorten the observable lifetime of minority carriers. \cite{Giorgioni14} Therefore, time-resolved PL offers us a very sensitive probe of the recombination dynamics and provides a direct method to identify the carrier loss mechanisms.

The samples were placed inside a closed-cycle cryostat at a temperature of 14 K and were excited by the 1064 nm line of a Nd:YAG Q-switched laser. The repetition frequency was 10 kHz with a temporal width of the laser pulse of about 10 ns and an estimated mean power density on the sample surface ranging from 1 to 22.5 $\mathrm{W\cdot cm^{-2}}$. The PL was collected by a single grating monochromator with a spectral bandpass set to 1.5 meV at 0.737 eV, which corresponds to the highest energy peak of the indirect gap PL band, as clarified by low temperature continuous-wave PL data of the as-grown sample shown in the inset of Fig.~\ref{fig2}. The emission was detected by a photomultiplier tube (PMT) operated in a single photon counting mode with a precision for the time decay of 0.3 ns. It should be noted that the weakening of the PL signal and the spectral position of the emission energy, being close to the detectivity cut-off of the PMT, prevented us from reliably implement time decay measurements above cryogenic temperatures.

A well-defined exponential decay can be observed in Fig.~\ref{fig2}(a). Surprisingly, the decay curves for both the as-grown and passivated samples demonstrate a similar slope, that is a comparable lifetime for electrons residing in the $L$-valley. The measured decay times, summarized in Fig.~\ref{fig2}(b), approach 30 ns and decrease by increasing the excitation power density, as a result of the local heating of the sample caused by the pulsed excitation. The negligible effect of the oxide coating on the carrier lifetime clearly demonstrates that, despite the large surface-to-volume ratio of the Ge micro-crystals, surface traps are not likely to limit the low temperature emission. This observation suggests the existence of competing mechanisms that are more effective than surface states in the capture of the injected carriers. 

To provide insight into the origin of such parasitic recombination, we introduce a sample that mimics the un-passivated, as-grown Ge micro-crystals except it encapsulates three additional $\mathrm{Si_{0.25}Ge_{0.75}}$ films at 2, 4 and 6 $\mu$m from the top surface, each with a thickness of 10 nm. The strain and Ge molar fractions have been suitably designed in order to provide a band off-set between the alloy layers and the Ge matrix that impedes the diffusion of the photogenerated carriers towards the defective Ge/Si interface. \cite{Pezzoli14} 

Remarkably, the insertion of the SiGe barriers yields the slowest decay among those reported in Fig.~\ref{fig2}(a) lengthening the electron lifetime up to $\approx 45$ ns [see Fig.~\ref{fig2}(b)]. This exceeds the literature data reported for Ge on Si heterostructures, \cite{Sukhdeo16, Sheng13, Geiger14, Nam14} and provides the first direct proof put forward in any recent work,\cite{Pezzoli14} that band-gap engineered SiGe/Ge architectures are effective in mitigating the optical activity of buried dislocations. 

At higher lattice temperatures, dislocations are nevertheless expected to play a less prominent role because of the augmented ionization probability of their associated localized energy levels. To clarify this physical picture we carried out additional PL measurements by using the 1.165 eV line of a CW Nd:$\mathrm{YVO_4}$ laser. The exciting power density was about 1 $\mathrm{kW / cm^2}$ and the emission was analyzed using a Fourier transform spectrometer equipped with Peltier cooled PbS detector. 

\begin{figure}
\includegraphics[width=8cm]{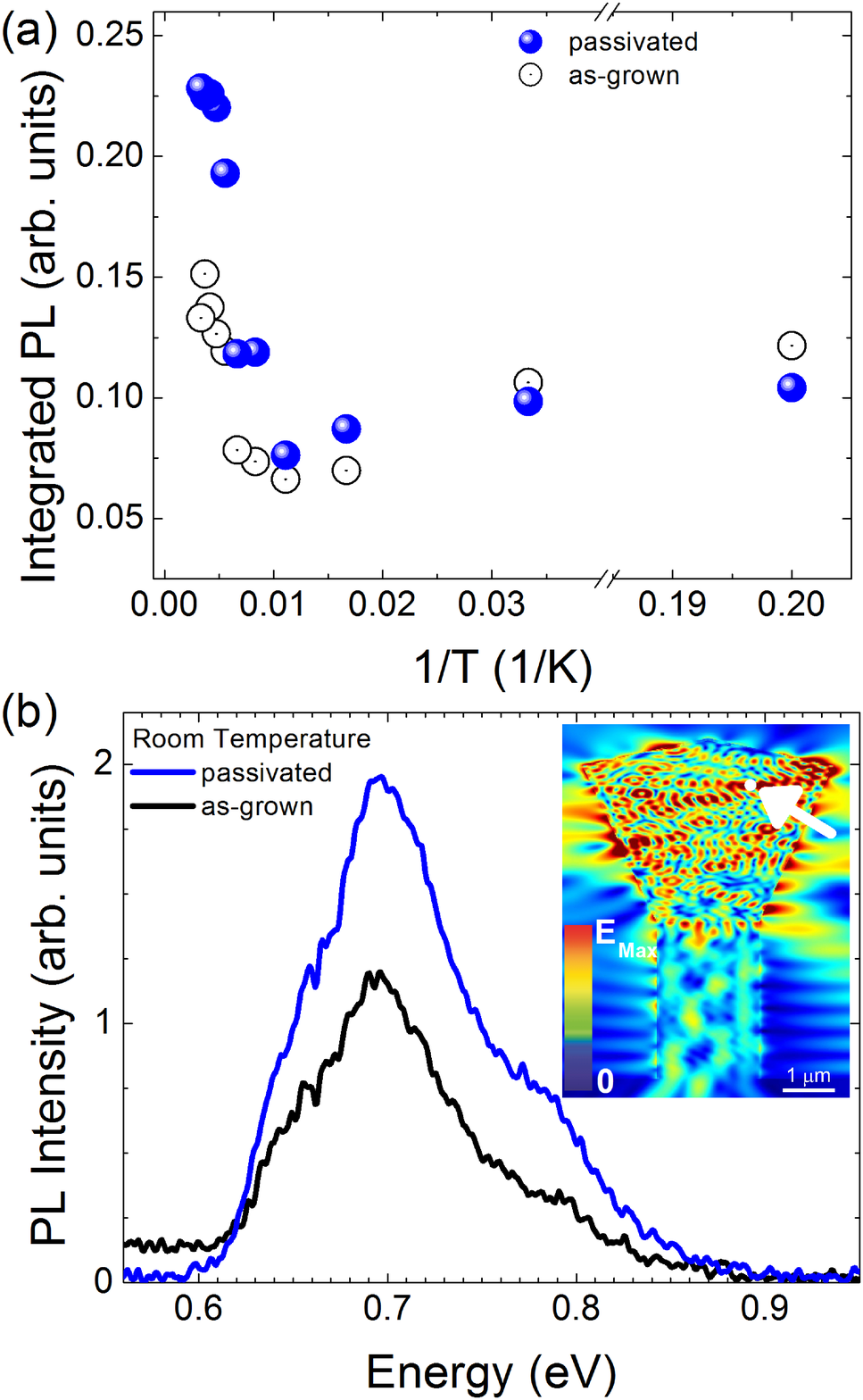}
 \caption{(a) The temperature dependence of the integrated photoluminescence (PL) intensity of the steady-state interband emission (direct plus indirect) for the as-grown (black dots) and the passivated (blue dots) samples. (b) The room temperature PL spectra of the as-grown (black line) and passivated (blue line) Ge micro-crystals under continuous wave excitation at 1.165 eV. The inset shows a cross-sectional map of the electric field distribution obtained by finite difference time domain calculations for an emitting point dipole placed inside a Ge micro-crystal at the position of the white circle indicated by the arrow.}
	\label{fig3}
 \end{figure}

Figure~\ref{fig3}(a) reports a notable deviation from the standard temperature dependent PL quenching. As the lattice temperature increases, there exists a sharp threshold above which a marked strengthening of the emission for both the as-grown and passivated sample can be observed. Such PL behavior is a hallmark of valley repopulation induced by thermal emission of carriers from the defect sites.\cite{Pezzoli14, Figielski78}  It should be noted, however, that the passivation results in a strong increase in the high temperature regime, yielding a room temperature PL spectrum with nearly twice the emission intensity of the as-grown Ge micro-crystals. This result demonstrates the effectiveness of the oxide coating in mitigating the nonradiative capture by surface states and it reconciles the optical investigation with the results of the electrical characterization discussed above.

Finally, in order to study the possible role of the oxide as an anti-reflection coating and to clarify whether the room temperature PL enhancement is affected by photonic effects, e.g. Purcell’s effect, we benchmarked our experimental data against finite-difference time-domain simulations.\cite{Pezzoli14,lumerical,Celebrano15} The steady-state PL process is mimicked by combining three simulation steps: (i) absorption of light by the Ge micro-crystals at the pump energy; (ii) local incoherent emission due to the recombination of electron-hole dipoles; and (iii) propagation of the emitted photons to the far field and their detection within the collection angle of the optics.\cite{Celebrano15} For the first step, we simulated the illumination of the Ge micro-crystals array with a Gaussian beam at 1.165 eV with an incidence angle resembling the actual experimental geometry. We then numerically evaluated the net flux of the Poynting vector entering the Ge micro-crystals and determined the total power absorption. As a second step, in order to mimic the incoherent emission process, we ran about 100 independent simulations, each one with a different randomly-oriented and randomly-located emitting dipole inside the Ge micro-crystals with a bandwidth covering the spectral range between 0.89 and 0.62 eV (i.e., both direct- and indirect-band emission from Ge). Such a large number of single-dipole simulations is needed in order to avoid spurious interference effects between different emitting dipoles, something that would have no counterpart in the actual PL process due to the lack of coherence. One representative field map resulting from such simulations is demonstrated in the inset of Fig.~\ref{fig3}(b). Finally, we projected the local field distribution generated by each dipole to the far field and integrated the power flux within the collection angle of the optics to get an estimate of the emitted power. 

By doing so, we are in the position to calculate the ratio between the expected PL intensity collected from the as-grown and passivated Ge micro-crystals under the same illumination conditions. We find that the difference in the PL intensity between the two systems is within 1-2\%, i.e. well below the experimentally observed effects. This confirms that the PL enhancement demonstrated in Fig.~\ref{fig3}(b) must be ascribed to an improvement of the internal quantum efficiency, which stems from the reduced nonradiative recombination at the passivated sidewalls of the micro-crystals.

In conclusion, we identified a viable approach to disentangle nonradiative transitions due to dislocations and surface states and gather insights about the complex kinetics of the recombination processes in Ge on Si heterostructures. We demonstrated the prominent role of dislocations at low temperature and the surge of room temperature PL emission under a suitable surface passivation. These findings can be used for engineering defects in key photonic building blocks such as lasers, optical resonator, and photodetectors fabricated in group IV materials. It is worth noting that our investigation can be extended to the epitaxial growth directly on silicon of other highly mismatched materials, thus enlarging even further the Si photonic toolbox.


%
%

%

We are grateful to the technical staff at the LNESS laboratory in Como and FIRST clean room at ETH Zuerich. We acknowledge support from Fondazione Cariplo through Grant No. 2013-0623, the Sinergia project NOVIPIX CRSII2\_147639 of the Swiss National Science Foundation and Pilegrowth Tech srl for technical support. 


\begin{thebibliography}{32}%
\makeatletter
\providecommand \@ifxundefined [1]{%
 \@ifx{#1\undefined}
}%
\providecommand \@ifnum [1]{%
 \ifnum #1\expandafter \@firstoftwo
 \else \expandafter \@secondoftwo
 \fi
}%
\providecommand \@ifx [1]{%
 \ifx #1\expandafter \@firstoftwo
 \else \expandafter \@secondoftwo
 \fi
}%
\providecommand \natexlab [1]{#1}%
\providecommand \enquote  [1]{``#1''}%
\providecommand \bibnamefont  [1]{#1}%
\providecommand \bibfnamefont [1]{#1}%
\providecommand \citenamefont [1]{#1}%
\providecommand \href@noop [0]{\@secondoftwo}%
\providecommand \href [0]{\begingroup \@sanitize@url \@href}%
\providecommand \@href[1]{\@@startlink{#1}\@@href}%
\providecommand \@@href[1]{\endgroup#1\@@endlink}%
\providecommand \@sanitize@url [0]{\catcode `\\12\catcode `\$12\catcode
  `\&12\catcode `\#12\catcode `\^12\catcode `\_12\catcode `\%12\relax}%
\providecommand \@@startlink[1]{}%
\providecommand \@@endlink[0]{}%
\providecommand \url  [0]{\begingroup\@sanitize@url \@url }%
\providecommand \@url [1]{\endgroup\@href {#1}{\urlprefix }}%
\providecommand \urlprefix  [0]{URL }%
\providecommand \Eprint [0]{\href }%
\providecommand \doibase [0]{http://dx.doi.org/}%
\providecommand \selectlanguage [0]{\@gobble}%
\providecommand \bibinfo  [0]{\@secondoftwo}%
\providecommand \bibfield  [0]{\@secondoftwo}%
\providecommand \translation [1]{[#1]}%
\providecommand \BibitemOpen [0]{}%
\providecommand \bibitemStop [0]{}%
\providecommand \bibitemNoStop [0]{.\EOS\space}%
\providecommand \EOS [0]{\spacefactor3000\relax}%
\providecommand \BibitemShut  [1]{\csname bibitem#1\endcsname}%
\let\auto@bib@innerbib\@empty
\bibitem [{\citenamefont {Leuthold}\ \emph {et~al.}(2010)\citenamefont
  {Leuthold}, \citenamefont {Koos},\ and\ \citenamefont {Freude}}]{Leuthold10}%
  \BibitemOpen
  \bibfield  {author} {\bibinfo {author} {\bibfnamefont {J.}~\bibnamefont
  {Leuthold}}, \bibinfo {author} {\bibfnamefont {C.}~\bibnamefont {Koos}}, \
  and\ \bibinfo {author} {\bibfnamefont {W.}~\bibnamefont {Freude}},\
  }\href@noop {} {\bibfield  {journal} {\bibinfo  {journal} {Nature Photon.}\
  }\textbf {\bibinfo {volume} {4}},\ \bibinfo {pages} {535} (\bibinfo {year}
  {2010})}\BibitemShut {NoStop}%
\bibitem [{\citenamefont {Dai}\ \emph {et~al.}(2012)\citenamefont {Dai},
  \citenamefont {Bauters},\ and\ \citenamefont {Bowers}}]{Dai12}%
  \BibitemOpen
  \bibfield  {author} {\bibinfo {author} {\bibfnamefont {D.}~\bibnamefont
  {Dai}}, \bibinfo {author} {\bibfnamefont {J.}~\bibnamefont {Bauters}}, \ and\
  \bibinfo {author} {\bibfnamefont {J.~E.}\ \bibnamefont {Bowers}},\
  }\href@noop {} {\bibfield  {journal} {\bibinfo  {journal} {Light: Science \&
  Applications}\ }\textbf {\bibinfo {volume} {1}},\ \bibinfo {pages} {e1}
  (\bibinfo {year} {2012})}\BibitemShut {NoStop}%
\bibitem [{\citenamefont {Lim}\ \emph {et~al.}(2014)\citenamefont {Lim},
  \citenamefont {Song}, \citenamefont {Fang}, \citenamefont {Li}, \citenamefont
  {Tu}, \citenamefont {Duan}, \citenamefont {Chen}, \citenamefont {Tern},\ and\
  \citenamefont {Liow}}]{Lim14}%
  \BibitemOpen
  \bibfield  {author} {\bibinfo {author} {\bibfnamefont {A.~E.-J.}\
  \bibnamefont {Lim}}, \bibinfo {author} {\bibfnamefont {J.}~\bibnamefont
  {Song}}, \bibinfo {author} {\bibfnamefont {Q.}~\bibnamefont {Fang}}, \bibinfo
  {author} {\bibfnamefont {C.}~\bibnamefont {Li}}, \bibinfo {author}
  {\bibfnamefont {X.}~\bibnamefont {Tu}}, \bibinfo {author} {\bibfnamefont
  {N.}~\bibnamefont {Duan}}, \bibinfo {author} {\bibfnamefont {K.~K.}\
  \bibnamefont {Chen}}, \bibinfo {author} {\bibfnamefont {R.~P.-C.}\
  \bibnamefont {Tern}}, \ and\ \bibinfo {author} {\bibfnamefont {T.-Y.}\
  \bibnamefont {Liow}},\ }\href@noop {} {\bibfield  {journal} {\bibinfo
  {journal} {IEEE J. Sel. Top. Quant.}\ }\textbf {\bibinfo {volume} {20}},\
  \bibinfo {pages} {8300112} (\bibinfo {year} {2014})}\BibitemShut {NoStop}%
\bibitem [{\citenamefont {Pillarisetty}(2011)}]{Pillarisetty11}%
  \BibitemOpen
  \bibfield  {author} {\bibinfo {author} {\bibfnamefont {R.}~\bibnamefont
  {Pillarisetty}},\ }\href@noop {} {\bibfield  {journal} {\bibinfo  {journal}
  {Nature}\ }\textbf {\bibinfo {volume} {479}},\ \bibinfo {pages} {324}
  (\bibinfo {year} {2011})}\BibitemShut {NoStop}%
\bibitem [{\citenamefont {Pezzoli}\ \emph {et~al.}(2013)\citenamefont
  {Pezzoli}, \citenamefont {Qing}, \citenamefont {Giorgioni}, \citenamefont
  {Isella}, \citenamefont {Grilli}, \citenamefont {Guzzi},\ and\ \citenamefont
  {Dery}}]{Pezzoli13}%
  \BibitemOpen
  \bibfield  {author} {\bibinfo {author} {\bibfnamefont {F.}~\bibnamefont
  {Pezzoli}}, \bibinfo {author} {\bibfnamefont {L.}~\bibnamefont {Qing}},
  \bibinfo {author} {\bibfnamefont {A.}~\bibnamefont {Giorgioni}}, \bibinfo
  {author} {\bibfnamefont {G.}~\bibnamefont {Isella}}, \bibinfo {author}
  {\bibfnamefont {E.}~\bibnamefont {Grilli}}, \bibinfo {author} {\bibfnamefont
  {M.}~\bibnamefont {Guzzi}}, \ and\ \bibinfo {author} {\bibfnamefont
  {H.}~\bibnamefont {Dery}},\ }\href@noop {} {\bibfield  {journal} {\bibinfo
  {journal} {Phys. Rev. B}\ }\textbf {\bibinfo {volume} {88}},\ \bibinfo
  {pages} {045204} (\bibinfo {year} {2013})}\BibitemShut {NoStop}%
\bibitem [{\citenamefont {Michel}\ \emph {et~al.}(2010)\citenamefont {Michel},
  \citenamefont {Liu},\ and\ \citenamefont {Kimerling}}]{Michel10}%
  \BibitemOpen
  \bibfield  {author} {\bibinfo {author} {\bibfnamefont {J.}~\bibnamefont
  {Michel}}, \bibinfo {author} {\bibfnamefont {J.~F.}\ \bibnamefont {Liu}}, \
  and\ \bibinfo {author} {\bibfnamefont {L.~C.}\ \bibnamefont {Kimerling}},\
  }\href@noop {} {\bibfield  {journal} {\bibinfo  {journal} {Nature Photon.}\
  }\textbf {\bibinfo {volume} {4}},\ \bibinfo {pages} {527} (\bibinfo {year}
  {2010})}\BibitemShut {NoStop}%
\bibitem [{\citenamefont {Liang}\ and\ \citenamefont {Bowers}(2010)}]{Liang10}%
  \BibitemOpen
  \bibfield  {author} {\bibinfo {author} {\bibfnamefont {D.}~\bibnamefont
  {Liang}}\ and\ \bibinfo {author} {\bibfnamefont {J.~E.}\ \bibnamefont
  {Bowers}},\ }\href@noop {} {\bibfield  {journal} {\bibinfo  {journal} {Nature
  Photon.}\ }\textbf {\bibinfo {volume} {4}},\ \bibinfo {pages} {511} (\bibinfo
  {year} {2010})}\BibitemShut {NoStop}%
\bibitem [{\citenamefont {Baldassarre}\ \emph {et~al.}(2015)\citenamefont
  {Baldassarre}, \citenamefont {Sakat}, \citenamefont {Frigerio}, \citenamefont
  {Samarelli}, \citenamefont {Gallacher}, \citenamefont {Calandrini},
  \citenamefont {Isella}, \citenamefont {Paul}, \citenamefont {Ortolani},\ and\
  \citenamefont {Biagioni}}]{Baldassarre15}%
  \BibitemOpen
  \bibfield  {author} {\bibinfo {author} {\bibfnamefont {L.}~\bibnamefont
  {Baldassarre}}, \bibinfo {author} {\bibfnamefont {E.}~\bibnamefont {Sakat}},
  \bibinfo {author} {\bibfnamefont {J.}~\bibnamefont {Frigerio}}, \bibinfo
  {author} {\bibfnamefont {A.}~\bibnamefont {Samarelli}}, \bibinfo {author}
  {\bibfnamefont {K.}~\bibnamefont {Gallacher}}, \bibinfo {author}
  {\bibfnamefont {E.}~\bibnamefont {Calandrini}}, \bibinfo {author}
  {\bibfnamefont {G.}~\bibnamefont {Isella}}, \bibinfo {author} {\bibfnamefont
  {D.~J.}\ \bibnamefont {Paul}}, \bibinfo {author} {\bibfnamefont
  {M.}~\bibnamefont {Ortolani}}, \ and\ \bibinfo {author} {\bibfnamefont
  {P.}~\bibnamefont {Biagioni}},\ }\href@noop {} {\bibfield  {journal}
  {\bibinfo  {journal} {Nano Lett.}\ }\textbf {\bibinfo {volume} {15}},\
  \bibinfo {pages} {7225} (\bibinfo {year} {2015})}\BibitemShut {NoStop}%
\bibitem [{\citenamefont {Paul}(2004)}]{Paul04}%
  \BibitemOpen
  \bibfield  {author} {\bibinfo {author} {\bibfnamefont {D.~J.}\ \bibnamefont
  {Paul}},\ }\href@noop {} {\bibfield  {journal} {\bibinfo  {journal}
  {Semicond. Sci. Technol.}\ }\textbf {\bibinfo {volume} {19}},\ \bibinfo
  {pages} {R75} (\bibinfo {year} {2004})}\BibitemShut {NoStop}%
\bibitem [{\citenamefont {Liu}\ \emph {et~al.}(2010)\citenamefont {Liu},
  \citenamefont {Sun}, \citenamefont {Camacho-Aguilera}, \citenamefont
  {Kimerling},\ and\ \citenamefont {Michel}}]{Liu10}%
  \BibitemOpen
  \bibfield  {author} {\bibinfo {author} {\bibfnamefont {J.}~\bibnamefont
  {Liu}}, \bibinfo {author} {\bibfnamefont {X.}~\bibnamefont {Sun}}, \bibinfo
  {author} {\bibfnamefont {R.}~\bibnamefont {Camacho-Aguilera}}, \bibinfo
  {author} {\bibfnamefont {L.~C.}\ \bibnamefont {Kimerling}}, \ and\ \bibinfo
  {author} {\bibfnamefont {J.}~\bibnamefont {Michel}},\ }\href@noop {}
  {\bibfield  {journal} {\bibinfo  {journal} {Opt. Lett.}\ }\textbf {\bibinfo
  {volume} {35}},\ \bibinfo {pages} {679} (\bibinfo {year} {2010})}\BibitemShut
  {NoStop}%
\bibitem [{\citenamefont {Camacho-Aguilera}\ \emph {et~al.}(2012)\citenamefont
  {Camacho-Aguilera}, \citenamefont {Cai}, \citenamefont {Patel}, \citenamefont
  {Bessette}, \citenamefont {Romagnoli}, \citenamefont {Kimerling},\ and\
  \citenamefont {Michel}}]{Camacho12}%
  \BibitemOpen
  \bibfield  {author} {\bibinfo {author} {\bibfnamefont {R.~E.}\ \bibnamefont
  {Camacho-Aguilera}}, \bibinfo {author} {\bibfnamefont {Y.}~\bibnamefont
  {Cai}}, \bibinfo {author} {\bibfnamefont {N.}~\bibnamefont {Patel}}, \bibinfo
  {author} {\bibfnamefont {J.~T.}\ \bibnamefont {Bessette}}, \bibinfo {author}
  {\bibfnamefont {M.}~\bibnamefont {Romagnoli}}, \bibinfo {author}
  {\bibfnamefont {L.~C.}\ \bibnamefont {Kimerling}}, \ and\ \bibinfo {author}
  {\bibfnamefont {J.}~\bibnamefont {Michel}},\ }\href@noop {} {\bibfield
  {journal} {\bibinfo  {journal} {Opt. Express}\ }\textbf {\bibinfo {volume}
  {20}},\ \bibinfo {pages} {11316} (\bibinfo {year} {2012})}\BibitemShut
  {NoStop}%
\bibitem [{\citenamefont {Wirths}\ \emph {et~al.}(2015)\citenamefont {Wirths},
  \citenamefont {Geiger}, \citenamefont {von~den Driesch}, \citenamefont
  {Mussler}, \citenamefont {Stoica}, \citenamefont {S.Mantl}, \citenamefont
  {Ikonic}, \citenamefont {Luysberg}, \citenamefont {Chiussi}, \citenamefont
  {Hartmann}, \citenamefont {Sigg}, \citenamefont {Faist}, \citenamefont
  {Buca},\ and\ \citenamefont {Gr{\"u}tzmacher}}]{Wirths15}%
  \BibitemOpen
  \bibfield  {author} {\bibinfo {author} {\bibfnamefont {S.}~\bibnamefont
  {Wirths}}, \bibinfo {author} {\bibfnamefont {R.}~\bibnamefont {Geiger}},
  \bibinfo {author} {\bibfnamefont {N.}~\bibnamefont {von~den Driesch}},
  \bibinfo {author} {\bibfnamefont {G.}~\bibnamefont {Mussler}}, \bibinfo
  {author} {\bibfnamefont {T.}~\bibnamefont {Stoica}}, \bibinfo {author}
  {\bibnamefont {S.Mantl}}, \bibinfo {author} {\bibfnamefont {Z.}~\bibnamefont
  {Ikonic}}, \bibinfo {author} {\bibfnamefont {M.}~\bibnamefont {Luysberg}},
  \bibinfo {author} {\bibfnamefont {S.}~\bibnamefont {Chiussi}}, \bibinfo
  {author} {\bibfnamefont {J.~M.}\ \bibnamefont {Hartmann}}, \bibinfo {author}
  {\bibfnamefont {H.}~\bibnamefont {Sigg}}, \bibinfo {author} {\bibfnamefont
  {J.}~\bibnamefont {Faist}}, \bibinfo {author} {\bibfnamefont
  {D.}~\bibnamefont {Buca}}, \ and\ \bibinfo {author} {\bibfnamefont
  {D.}~\bibnamefont {Gr{\"u}tzmacher}},\ }\href@noop {} {\bibfield  {journal}
  {\bibinfo  {journal} {Nature Photon.}\ }\textbf {\bibinfo {volume} {9}},\
  \bibinfo {pages} {88} (\bibinfo {year} {2015})}\BibitemShut {NoStop}%
\bibitem [{\citenamefont {Yang}\ \emph {et~al.}(2003)\citenamefont {Yang},
  \citenamefont {Groenert}, \citenamefont {Leitz}, \citenamefont {Pitera},
  \citenamefont {Currie},\ and\ \citenamefont {Fitzgerald}}]{Yang03}%
  \BibitemOpen
  \bibfield  {author} {\bibinfo {author} {\bibfnamefont {V.~K.}\ \bibnamefont
  {Yang}}, \bibinfo {author} {\bibfnamefont {M.}~\bibnamefont {Groenert}},
  \bibinfo {author} {\bibfnamefont {C.~W.}\ \bibnamefont {Leitz}}, \bibinfo
  {author} {\bibfnamefont {A.~J.}\ \bibnamefont {Pitera}}, \bibinfo {author}
  {\bibfnamefont {M.~T.}\ \bibnamefont {Currie}}, \ and\ \bibinfo {author}
  {\bibfnamefont {E.~A.}\ \bibnamefont {Fitzgerald}},\ }\href@noop {}
  {\bibfield  {journal} {\bibinfo  {journal} {J. Appl. Phys.}\ }\textbf
  {\bibinfo {volume} {93}},\ \bibinfo {pages} {3859} (\bibinfo {year}
  {2003})}\BibitemShut {NoStop}%
\bibitem [{\citenamefont {Sukhdeo}\ \emph {et~al.}(2016)\citenamefont
  {Sukhdeo}, \citenamefont {Gupta}, \citenamefont {Saraswat}, \citenamefont
  {Dutt},\ and\ \citenamefont {Nam}}]{Sukhdeo16}%
  \BibitemOpen
  \bibfield  {author} {\bibinfo {author} {\bibfnamefont {D.~S.}\ \bibnamefont
  {Sukhdeo}}, \bibinfo {author} {\bibfnamefont {S.}~\bibnamefont {Gupta}},
  \bibinfo {author} {\bibfnamefont {K.~C.}\ \bibnamefont {Saraswat}}, \bibinfo
  {author} {\bibfnamefont {B.~R.}\ \bibnamefont {Dutt}}, \ and\ \bibinfo
  {author} {\bibfnamefont {D.}~\bibnamefont {Nam}},\ }\href@noop {} {\bibfield
  {journal} {\bibinfo  {journal} {Opt. Commun.}\ }\textbf {\bibinfo {volume}
  {364}},\ \bibinfo {pages} {233} (\bibinfo {year} {2016})}\BibitemShut
  {NoStop}%
\bibitem [{\citenamefont {Li}\ \emph {et~al.}()\citenamefont {Li},
  \citenamefont {Li}, \citenamefont {Li}, \citenamefont {Chrostowski},\ and\
  \citenamefont {Xia}}]{Li16}%
  \BibitemOpen
  \bibfield  {author} {\bibinfo {author} {\bibfnamefont {X.}~\bibnamefont
  {Li}}, \bibinfo {author} {\bibfnamefont {Z.}~\bibnamefont {Li}}, \bibinfo
  {author} {\bibfnamefont {S.}~\bibnamefont {Li}}, \bibinfo {author}
  {\bibfnamefont {L.}~\bibnamefont {Chrostowski}}, \ and\ \bibinfo {author}
  {\bibfnamefont {G.}~\bibnamefont {Xia}},\ }\href@noop {} {\ }\Eprint
  {http://arxiv.org/abs/1511.05972} {arXiv:1511.05972} \BibitemShut {NoStop}%
\bibitem [{\citenamefont {Falub}\ \emph {et~al.}(2012)\citenamefont {Falub},
  \citenamefont {von K{\"a}nel}, \citenamefont {Isa}, \citenamefont
  {Bergamaschini}, \citenamefont {Marzegalli}, \citenamefont {Chrastina},
  \citenamefont {Isella}, \citenamefont {M{\"u}ller}, \citenamefont
  {Niedermann},\ and\ \citenamefont {Miglio}}]{Falub2012}%
  \BibitemOpen
  \bibfield  {author} {\bibinfo {author} {\bibfnamefont {C.~V.}\ \bibnamefont
  {Falub}}, \bibinfo {author} {\bibfnamefont {H.}~\bibnamefont {von
  K{\"a}nel}}, \bibinfo {author} {\bibfnamefont {F.}~\bibnamefont {Isa}},
  \bibinfo {author} {\bibfnamefont {R.}~\bibnamefont {Bergamaschini}}, \bibinfo
  {author} {\bibfnamefont {A.}~\bibnamefont {Marzegalli}}, \bibinfo {author}
  {\bibfnamefont {D.}~\bibnamefont {Chrastina}}, \bibinfo {author}
  {\bibfnamefont {G.}~\bibnamefont {Isella}}, \bibinfo {author} {\bibfnamefont
  {E.}~\bibnamefont {M{\"u}ller}}, \bibinfo {author} {\bibfnamefont
  {P.}~\bibnamefont {Niedermann}}, \ and\ \bibinfo {author} {\bibfnamefont
  {L.}~\bibnamefont {Miglio}},\ }\href@noop {} {\bibfield  {journal} {\bibinfo
  {journal} {Science}\ }\textbf {\bibinfo {volume} {335}},\ \bibinfo {pages}
  {6074} (\bibinfo {year} {2012})}\BibitemShut {NoStop}%
\bibitem [{\citenamefont {Rosenblad}\ \emph {et~al.}(1998)\citenamefont
  {Rosenblad}, \citenamefont {Deller}, \citenamefont {Dommann}, \citenamefont
  {Meyer}, \citenamefont {Schr{\"o}ter},\ and\ \citenamefont {von
  K{\"a}nel}}]{Rosenblad98}%
  \BibitemOpen
  \bibfield  {author} {\bibinfo {author} {\bibfnamefont {C.}~\bibnamefont
  {Rosenblad}}, \bibinfo {author} {\bibfnamefont {H.~R.}\ \bibnamefont
  {Deller}}, \bibinfo {author} {\bibfnamefont {A.}~\bibnamefont {Dommann}},
  \bibinfo {author} {\bibfnamefont {T.}~\bibnamefont {Meyer}}, \bibinfo
  {author} {\bibfnamefont {P.}~\bibnamefont {Schr{\"o}ter}}, \ and\ \bibinfo
  {author} {\bibfnamefont {H.}~\bibnamefont {von K{\"a}nel}},\ }\href@noop {}
  {\bibfield  {journal} {\bibinfo  {journal} {J. Vac. Sci. Technol. A}\
  }\textbf {\bibinfo {volume} {16}},\ \bibinfo {pages} {2785} (\bibinfo {year}
  {1998})}\BibitemShut {NoStop}%
\bibitem [{\citenamefont {Marzegalli}\ \emph {et~al.}(2013)\citenamefont
  {Marzegalli}, \citenamefont {Isa}, \citenamefont {Groiss}, \citenamefont
  {M{\"u}ller}, \citenamefont {Falub}, \citenamefont {Taboada}, \citenamefont
  {Niedermann}, \citenamefont {Isella}, \citenamefont {Sch{\"a}ffler},
  \citenamefont {Montalenti}, \citenamefont {von K{\"a}nel},\ and\
  \citenamefont {Leo}}]{Marzegalli13}%
  \BibitemOpen
  \bibfield  {author} {\bibinfo {author} {\bibfnamefont {A.}~\bibnamefont
  {Marzegalli}}, \bibinfo {author} {\bibfnamefont {F.}~\bibnamefont {Isa}},
  \bibinfo {author} {\bibfnamefont {H.}~\bibnamefont {Groiss}}, \bibinfo
  {author} {\bibfnamefont {E.}~\bibnamefont {M{\"u}ller}}, \bibinfo {author}
  {\bibfnamefont {C.~V.}\ \bibnamefont {Falub}}, \bibinfo {author}
  {\bibfnamefont {A.~G.}\ \bibnamefont {Taboada}}, \bibinfo {author}
  {\bibfnamefont {P.}~\bibnamefont {Niedermann}}, \bibinfo {author}
  {\bibfnamefont {G.}~\bibnamefont {Isella}}, \bibinfo {author} {\bibfnamefont
  {F.}~\bibnamefont {Sch{\"a}ffler}}, \bibinfo {author} {\bibfnamefont
  {F.}~\bibnamefont {Montalenti}}, \bibinfo {author} {\bibfnamefont
  {H.}~\bibnamefont {von K{\"a}nel}}, \ and\ \bibinfo {author} {\bibfnamefont
  {M.}~\bibnamefont {Leo}},\ }\href@noop {} {\bibfield  {journal} {\bibinfo
  {journal} {Adv. Mater.}\ }\textbf {\bibinfo {volume} {25}},\ \bibinfo {pages}
  {4408} (\bibinfo {year} {2013})}\BibitemShut {NoStop}%
\bibitem [{\citenamefont {Isa}\ \emph {et~al.}(2013)\citenamefont {Isa},
  \citenamefont {Marzegalli}, \citenamefont {Taboada}, \citenamefont {Falub},
  \citenamefont {Isella}, \citenamefont {Montalenti}, \citenamefont {von
  K{\"a}nel},\ and\ \citenamefont {Miglio}}]{Isa13}%
  \BibitemOpen
  \bibfield  {author} {\bibinfo {author} {\bibfnamefont {F.}~\bibnamefont
  {Isa}}, \bibinfo {author} {\bibfnamefont {A.}~\bibnamefont {Marzegalli}},
  \bibinfo {author} {\bibfnamefont {A.~G.}\ \bibnamefont {Taboada}}, \bibinfo
  {author} {\bibfnamefont {C.~V.}\ \bibnamefont {Falub}}, \bibinfo {author}
  {\bibfnamefont {G.}~\bibnamefont {Isella}}, \bibinfo {author} {\bibfnamefont
  {F.}~\bibnamefont {Montalenti}}, \bibinfo {author} {\bibfnamefont
  {H.}~\bibnamefont {von K{\"a}nel}}, \ and\ \bibinfo {author} {\bibfnamefont
  {L.}~\bibnamefont {Miglio}},\ }\href@noop {} {\bibfield  {journal} {\bibinfo
  {journal} {APL Materials}\ }\textbf {\bibinfo {volume} {1}},\ \bibinfo
  {pages} {052109} (\bibinfo {year} {2013})}\BibitemShut {NoStop}%
\bibitem [{\citenamefont {Pezzoli}\ \emph {et~al.}(2014)\citenamefont
  {Pezzoli}, \citenamefont {Isa}, \citenamefont {Isella}, \citenamefont
  {Falub}, \citenamefont {Kreiliger}, \citenamefont {Salvalaglio},
  \citenamefont {Bergamaschini}, \citenamefont {Grilli}, \citenamefont {Guzzi},
  \citenamefont {von K{\"a}nel},\ and\ \citenamefont {Miglio}}]{Pezzoli14}%
  \BibitemOpen
  \bibfield  {author} {\bibinfo {author} {\bibfnamefont {F.}~\bibnamefont
  {Pezzoli}}, \bibinfo {author} {\bibfnamefont {F.}~\bibnamefont {Isa}},
  \bibinfo {author} {\bibfnamefont {G.}~\bibnamefont {Isella}}, \bibinfo
  {author} {\bibfnamefont {C.~V.}\ \bibnamefont {Falub}}, \bibinfo {author}
  {\bibfnamefont {T.}~\bibnamefont {Kreiliger}}, \bibinfo {author}
  {\bibfnamefont {M.}~\bibnamefont {Salvalaglio}}, \bibinfo {author}
  {\bibfnamefont {R.}~\bibnamefont {Bergamaschini}}, \bibinfo {author}
  {\bibfnamefont {E.}~\bibnamefont {Grilli}}, \bibinfo {author} {\bibfnamefont
  {M.}~\bibnamefont {Guzzi}}, \bibinfo {author} {\bibfnamefont
  {H.}~\bibnamefont {von K{\"a}nel}}, \ and\ \bibinfo {author} {\bibfnamefont
  {L.}~\bibnamefont {Miglio}},\ }\href@noop {} {\bibfield  {journal} {\bibinfo
  {journal} {Phys. Rev. Appl.}\ }\textbf {\bibinfo {volume} {1}},\ \bibinfo
  {pages} {044005} (\bibinfo {year} {2014})}\BibitemShut {NoStop}%
\bibitem [{\citenamefont {Isa}\ \emph {et~al.}(2015)\citenamefont {Isa},
  \citenamefont {Pezzoli}, \citenamefont {Isella}, \citenamefont {Medu{\v n}a},
  \citenamefont {Falub}, \citenamefont {M{\"u}ller}, \citenamefont {Kreiliger},
  \citenamefont {Taboada}, \citenamefont {von K{\"a}nel},\ and\ \citenamefont
  {Miglio}}]{Isa15}%
  \BibitemOpen
  \bibfield  {author} {\bibinfo {author} {\bibfnamefont {F.}~\bibnamefont
  {Isa}}, \bibinfo {author} {\bibfnamefont {F.}~\bibnamefont {Pezzoli}},
  \bibinfo {author} {\bibfnamefont {G.}~\bibnamefont {Isella}}, \bibinfo
  {author} {\bibfnamefont {M.}~\bibnamefont {Medu{\v n}a}}, \bibinfo {author}
  {\bibfnamefont {C.~V.}\ \bibnamefont {Falub}}, \bibinfo {author}
  {\bibfnamefont {E.}~\bibnamefont {M{\"u}ller}}, \bibinfo {author}
  {\bibfnamefont {T.}~\bibnamefont {Kreiliger}}, \bibinfo {author}
  {\bibfnamefont {A.~G.}\ \bibnamefont {Taboada}}, \bibinfo {author}
  {\bibfnamefont {H.}~\bibnamefont {von K{\"a}nel}}, \ and\ \bibinfo {author}
  {\bibfnamefont {L.}~\bibnamefont {Miglio}},\ }\href@noop {} {\bibfield
  {journal} {\bibinfo  {journal} {Semicond. Sci. Technol.}\ }\textbf {\bibinfo
  {volume} {30}},\ \bibinfo {pages} {105001} (\bibinfo {year}
  {2015})}\BibitemShut {NoStop}%
\bibitem [{\citenamefont {Matsubara}\ \emph {et~al.}(2008)\citenamefont
  {Matsubara}, \citenamefont {Sasada}, \citenamefont {Takenaka},\ and\
  \citenamefont {Takagi}}]{Matsubara08}%
  \BibitemOpen
  \bibfield  {author} {\bibinfo {author} {\bibfnamefont {H.}~\bibnamefont
  {Matsubara}}, \bibinfo {author} {\bibfnamefont {T.}~\bibnamefont {Sasada}},
  \bibinfo {author} {\bibfnamefont {M.}~\bibnamefont {Takenaka}}, \ and\
  \bibinfo {author} {\bibfnamefont {S.}~\bibnamefont {Takagi}},\ }\href@noop {}
  {\bibfield  {journal} {\bibinfo  {journal} {Appl. Phys Lett.}\ }\textbf
  {\bibinfo {volume} {93}},\ \bibinfo {pages} {032104} (\bibinfo {year}
  {2008})}\BibitemShut {NoStop}%
\bibitem [{\citenamefont {Millar}\ \emph {et~al.}(2015)\citenamefont {Millar},
  \citenamefont {Gallacher}, \citenamefont {Samarelli}, \citenamefont
  {Frigerio}, \citenamefont {Chrastina}, \citenamefont {Isella}, \citenamefont
  {Dieing},\ and\ \citenamefont {Paul}}]{Millar15}%
  \BibitemOpen
  \bibfield  {author} {\bibinfo {author} {\bibfnamefont {R.~W.}\ \bibnamefont
  {Millar}}, \bibinfo {author} {\bibfnamefont {K.}~\bibnamefont {Gallacher}},
  \bibinfo {author} {\bibfnamefont {A.}~\bibnamefont {Samarelli}}, \bibinfo
  {author} {\bibfnamefont {J.}~\bibnamefont {Frigerio}}, \bibinfo {author}
  {\bibfnamefont {D.}~\bibnamefont {Chrastina}}, \bibinfo {author}
  {\bibfnamefont {G.}~\bibnamefont {Isella}}, \bibinfo {author} {\bibfnamefont
  {T.}~\bibnamefont {Dieing}}, \ and\ \bibinfo {author} {\bibfnamefont {D.~J.}\
  \bibnamefont {Paul}},\ }\href@noop {} {\bibfield  {journal} {\bibinfo
  {journal} {Opt. Express}\ }\textbf {\bibinfo {volume} {23}},\ \bibinfo
  {pages} {18193} (\bibinfo {year} {2015})}\BibitemShut {NoStop}%
\bibitem [{\citenamefont {Bellenger}\ \emph {et~al.}(2008)\citenamefont
  {Bellenger}, \citenamefont {Houssa}, \citenamefont {Delabie}, \citenamefont
  {Afanasiev}, \citenamefont {Conard}, \citenamefont {Caymax}, \citenamefont
  {Meuris}, \citenamefont {Meyer},\ and\ \citenamefont {Heyns}}]{Bellenger08}%
  \BibitemOpen
  \bibfield  {author} {\bibinfo {author} {\bibfnamefont {F.}~\bibnamefont
  {Bellenger}}, \bibinfo {author} {\bibfnamefont {M.}~\bibnamefont {Houssa}},
  \bibinfo {author} {\bibfnamefont {A.}~\bibnamefont {Delabie}}, \bibinfo
  {author} {\bibfnamefont {V.}~\bibnamefont {Afanasiev}}, \bibinfo {author}
  {\bibfnamefont {T.}~\bibnamefont {Conard}}, \bibinfo {author} {\bibfnamefont
  {M.}~\bibnamefont {Caymax}}, \bibinfo {author} {\bibfnamefont
  {M.}~\bibnamefont {Meuris}}, \bibinfo {author} {\bibfnamefont {K.~D.}\
  \bibnamefont {Meyer}}, \ and\ \bibinfo {author} {\bibfnamefont {M.~M.}\
  \bibnamefont {Heyns}},\ }\href@noop {} {\bibfield  {journal} {\bibinfo
  {journal} {J. Electrochem Soc.}\ }\textbf {\bibinfo {volume} {155}},\
  \bibinfo {pages} {G33} (\bibinfo {year} {2008})}\BibitemShut {NoStop}%
\bibitem [{\citenamefont {Hirayama}\ \emph {et~al.}(2011)\citenamefont
  {Hirayama}, \citenamefont {Yoshino}, \citenamefont {Ueno}, \citenamefont
  {Iwamura}, \citenamefont {Yang}, \citenamefont {Wang},\ and\ \citenamefont
  {Nakashima}}]{Hirayama11}%
  \BibitemOpen
  \bibfield  {author} {\bibinfo {author} {\bibfnamefont {K.}~\bibnamefont
  {Hirayama}}, \bibinfo {author} {\bibfnamefont {K.}~\bibnamefont {Yoshino}},
  \bibinfo {author} {\bibfnamefont {R.}~\bibnamefont {Ueno}}, \bibinfo {author}
  {\bibfnamefont {Y.}~\bibnamefont {Iwamura}}, \bibinfo {author} {\bibfnamefont
  {H.}~\bibnamefont {Yang}}, \bibinfo {author} {\bibfnamefont {D.}~\bibnamefont
  {Wang}}, \ and\ \bibinfo {author} {\bibfnamefont {H.}~\bibnamefont
  {Nakashima}},\ }\href@noop {} {\bibfield  {journal} {\bibinfo  {journal}
  {Solid State Electron.}\ }\textbf {\bibinfo {volume} {60}},\ \bibinfo {pages}
  {122} (\bibinfo {year} {2011})}\BibitemShut {NoStop}%
\bibitem [{\citenamefont {Giorgioni}\ \emph {et~al.}(2014)\citenamefont
  {Giorgioni}, \citenamefont {Vitiello}, \citenamefont {Grilli}, \citenamefont
  {Guzzi},\ and\ \citenamefont {Pezzoli}}]{Giorgioni14}%
  \BibitemOpen
  \bibfield  {author} {\bibinfo {author} {\bibfnamefont {A.}~\bibnamefont
  {Giorgioni}}, \bibinfo {author} {\bibfnamefont {E.}~\bibnamefont {Vitiello}},
  \bibinfo {author} {\bibfnamefont {E.}~\bibnamefont {Grilli}}, \bibinfo
  {author} {\bibfnamefont {M.}~\bibnamefont {Guzzi}}, \ and\ \bibinfo {author}
  {\bibfnamefont {F.}~\bibnamefont {Pezzoli}},\ }\href@noop {} {\bibfield
  {journal} {\bibinfo  {journal} {Appl. Phys. Lett.}\ }\textbf {\bibinfo
  {volume} {105}},\ \bibinfo {eid} {152404} (\bibinfo {year}
  {2014})}\BibitemShut {NoStop}%
\bibitem [{\citenamefont {Sheng}\ \emph {et~al.}(2013)\citenamefont {Sheng},
  \citenamefont {Leonhardt}, \citenamefont {Han}, \citenamefont {Johnston},
  \citenamefont {Cederberg},\ and\ \citenamefont {Carroll}}]{Sheng13}%
  \BibitemOpen
  \bibfield  {author} {\bibinfo {author} {\bibfnamefont {J.~J.}\ \bibnamefont
  {Sheng}}, \bibinfo {author} {\bibfnamefont {D.}~\bibnamefont {Leonhardt}},
  \bibinfo {author} {\bibfnamefont {S.~M.}\ \bibnamefont {Han}}, \bibinfo
  {author} {\bibfnamefont {S.~W.}\ \bibnamefont {Johnston}}, \bibinfo {author}
  {\bibfnamefont {J.~G.}\ \bibnamefont {Cederberg}}, \ and\ \bibinfo {author}
  {\bibfnamefont {M.~S.}\ \bibnamefont {Carroll}},\ }\href@noop {} {\bibfield
  {journal} {\bibinfo  {journal} {J. Vac. Sci. Technol. B}\ }\textbf {\bibinfo
  {volume} {31}},\ \bibinfo {pages} {051201} (\bibinfo {year}
  {2013})}\BibitemShut {NoStop}%
\bibitem [{\citenamefont {Geiger}\ \emph {et~al.}(2014)\citenamefont {Geiger},
  \citenamefont {Frigerio}, \citenamefont {S{\"u}ess}, \citenamefont
  {Chrastina}, \citenamefont {Isella}, \citenamefont {Spolenak}, \citenamefont
  {Faist},\ and\ \citenamefont {Sigg}}]{Geiger14}%
  \BibitemOpen
  \bibfield  {author} {\bibinfo {author} {\bibfnamefont {R.}~\bibnamefont
  {Geiger}}, \bibinfo {author} {\bibfnamefont {J.}~\bibnamefont {Frigerio}},
  \bibinfo {author} {\bibfnamefont {M.~J.}\ \bibnamefont {S{\"u}ess}}, \bibinfo
  {author} {\bibfnamefont {D.}~\bibnamefont {Chrastina}}, \bibinfo {author}
  {\bibfnamefont {G.}~\bibnamefont {Isella}}, \bibinfo {author} {\bibfnamefont
  {R.}~\bibnamefont {Spolenak}}, \bibinfo {author} {\bibfnamefont
  {J.}~\bibnamefont {Faist}}, \ and\ \bibinfo {author} {\bibfnamefont
  {H.}~\bibnamefont {Sigg}},\ }\href@noop {} {\bibfield  {journal} {\bibinfo
  {journal} {Appl. Phys. Lett.}\ }\textbf {\bibinfo {volume} {104}},\ \bibinfo
  {pages} {062106} (\bibinfo {year} {2014})}\BibitemShut {NoStop}%
\bibitem [{\citenamefont {Nam}\ \emph {et~al.}(2014)\citenamefont {Nam},
  \citenamefont {Kang}, \citenamefont {Brongersma},\ and\ \citenamefont
  {Saraswat}}]{Nam14}%
  \BibitemOpen
  \bibfield  {author} {\bibinfo {author} {\bibfnamefont {D.}~\bibnamefont
  {Nam}}, \bibinfo {author} {\bibfnamefont {J.~H.}\ \bibnamefont {Kang}},
  \bibinfo {author} {\bibfnamefont {M.~L.}\ \bibnamefont {Brongersma}}, \ and\
  \bibinfo {author} {\bibfnamefont {K.~C.}\ \bibnamefont {Saraswat}},\
  }\href@noop {} {\bibfield  {journal} {\bibinfo  {journal} {Opt. Lett.}\
  }\textbf {\bibinfo {volume} {39}},\ \bibinfo {pages} {6205} (\bibinfo {year}
  {2014})}\BibitemShut {NoStop}%
\bibitem [{\citenamefont {Figielski}(1978)}]{Figielski78}%
  \BibitemOpen
  \bibfield  {author} {\bibinfo {author} {\bibfnamefont {T.}~\bibnamefont
  {Figielski}},\ }\href@noop {} {\bibfield  {journal} {\bibinfo  {journal}
  {Solid-State Electron.}\ }\textbf {\bibinfo {volume} {21}},\ \bibinfo {pages}
  {1403} (\bibinfo {year} {1978})}\BibitemShut {NoStop}%
\bibitem [{FDTD Solutions, version 8.5.3; Lumerical Solutions, Inc.: Canada,
  2013()}]{lumerical}%
  \BibitemOpen
  FDTD Solutions, version 8.5.3; Lumerical Solutions, Inc.: Canada, 2013,\
  \href@noop {} {}\BibitemShut {NoStop}%
\bibitem [{\citenamefont {Celebrano}\ \emph {et~al.}(2015)\citenamefont
  {Celebrano}, \citenamefont {Baselli}, \citenamefont {Bollani}, \citenamefont
  {Frigerio}, \citenamefont {Shehata}, \citenamefont {Frera}, \citenamefont
  {Tosi}, \citenamefont {Farina}, \citenamefont {Pezzoli}, \citenamefont
  {Osmond}, \citenamefont {Wu}, \citenamefont {Hecht}, \citenamefont {Sordan},
  \citenamefont {Chrastina}, \citenamefont {Isella}, \citenamefont {Du{\'o}},
  \citenamefont {Finazzi},\ and\ \citenamefont {Biagioni}}]{Celebrano15}%
  \BibitemOpen
  \bibfield  {author} {\bibinfo {author} {\bibfnamefont {M.}~\bibnamefont
  {Celebrano}}, \bibinfo {author} {\bibfnamefont {M.}~\bibnamefont {Baselli}},
  \bibinfo {author} {\bibfnamefont {M.}~\bibnamefont {Bollani}}, \bibinfo
  {author} {\bibfnamefont {J.}~\bibnamefont {Frigerio}}, \bibinfo {author}
  {\bibfnamefont {A.~B.}\ \bibnamefont {Shehata}}, \bibinfo {author}
  {\bibfnamefont {A.~D.}\ \bibnamefont {Frera}}, \bibinfo {author}
  {\bibfnamefont {A.}~\bibnamefont {Tosi}}, \bibinfo {author} {\bibfnamefont
  {A.}~\bibnamefont {Farina}}, \bibinfo {author} {\bibfnamefont
  {F.}~\bibnamefont {Pezzoli}}, \bibinfo {author} {\bibfnamefont
  {J.}~\bibnamefont {Osmond}}, \bibinfo {author} {\bibfnamefont
  {X.}~\bibnamefont {Wu}}, \bibinfo {author} {\bibfnamefont {B.}~\bibnamefont
  {Hecht}}, \bibinfo {author} {\bibfnamefont {R.}~\bibnamefont {Sordan}},
  \bibinfo {author} {\bibfnamefont {D.}~\bibnamefont {Chrastina}}, \bibinfo
  {author} {\bibfnamefont {G.}~\bibnamefont {Isella}}, \bibinfo {author}
  {\bibfnamefont {L.}~\bibnamefont {Du{\'o}}}, \bibinfo {author} {\bibfnamefont
  {M.}~\bibnamefont {Finazzi}}, \ and\ \bibinfo {author} {\bibfnamefont
  {P.}~\bibnamefont {Biagioni}},\ }\href@noop {} {\bibfield  {journal}
  {\bibinfo  {journal} {ACS Photonics}\ }\textbf {\bibinfo {volume} {2}},\
  \bibinfo {pages} {53} (\bibinfo {year} {2015})}\BibitemShut {NoStop}%
\end{thebibliography}

%


\end{document}